\documentclass[aps,nofootinbib,twocolumn,prd]{revtex4}
\usepackage{ifpdf}
\usepackage{hyperref}
\usepackage{amsmath}
\usepackage{amssymb}
\usepackage{graphics}
\usepackage{natbib}
\usepackage{graphicx}
\usepackage{color}

\newcommand\hidetosubmit[1]{}
\renewcommand\hidetosubmit[1]{#1}
\newcommand\ForInternalReference[1]{}

\newcommand\unit[1]{\, {\rm #1}}
\newcommand\E[1]{\left<#1\right>}
\newcommand\mc{ {{\cal M}_c}}

\newcommand\Y[1]{Y^{(#1)}}

\newcommand\ST[1]{{\texttt{ST#1}}}
\newcommand\STLong[1]{{\texttt{SpinTaylor#1}}}

\newcommand\citeMCMC{\cite{LIGO-CBC-S6-PE,2011PhRvD..83h2002D,2011PhRvD..84f2003C,gr-extensions-tests-Europeans2011,gwastro-mergers-PE-Aylott-LIGOATest,2011ApJ...739...99N,2012PhRvD..85j4045V,gw-astro-PE-Raymond,gw-astro-PE-lalinference-v1}}

\def\bbh#1{binary black hole#1 (BBH#1)\gdef\bbh{BBH}}
\def\bh#1{black hole#1 (BH#1)\gdef\bh{BH}}

\begin{document}
\title{Rapid gravitational wave parameter estimation with a single spin: \\  Systematic uncertainties in parameter
  estimation with the SpinTaylorF2 approximation } 
\author{B. Miller}
\author{R. O'Shaughnessy}
\affiliation{Center for Computational Relativity and Gravitation, Rochester Institute of Technology, Rochester, NY 14623, USA}
\email{oshaughn@mail.rit.edu}
\author{T.B. Littenberg}
\affiliation{Center for Interdisciplinary Exploration and Research in Astrophysics (CIERA)\&
Dept. of Physics and Astronomy, 2145 Sheridan Rd, Evanston, IL 60208, USA}
\author{B. Farr}
\affiliation{Enrico Fermi Institute, University of Chicago, Chicago, IL 60637, USA}

\date{\today}
\begin{abstract}
Reliable low-latency gravitational wave parameter estimation is essential to target limited electromagnetic followup
facilities toward  astrophysically interesting and electromagnetically relevant sources of gravitational waves.  In this
study, we examine the tradeoff between speed and accuracy.  Specifically, we estimate the astrophysical relevance of systematic errors in the posterior parameter distributions
derived using   a fast-but-approximate waveform model, \STLong{F2} (\ST{F2}), in parameter estimation with \texttt{lalinference\_mcmc}.  
Though efficient, the \ST{F2} approximation to compact binary inspiral employs approximate kinematics
(e.g., a single spin) and an approximate waveform (e.g., frequency domain versus time domain).   More broadly, using a
large astrophysically-motivated population of generic compact binary merger signals, we
report on the effectualness and limitations of this single-spin approximation as a method to infer parameters of generic
compact binary sources.  
 For most low-mass compact binary sources, we find that the
\ST{F2} approximation estimates  compact binary parameters with
biases comparable to systematic uncertainties in the waveform.  We illustrate by example the effect these systematic
errors have on posterior probabilities most relevant to low-latency electromagnetic followup:  whether the secondary 
has a mass consistent with a neutron star; whether the masses, spins, and orbit are consistent with that neutron star's
tidal disruption; and whether the  binary's angular momentum axis is oriented along the line of sight.
\end{abstract}
\maketitle

\section{Introduction}
Ground based gravitational wave detector networks (notably advanced LIGO \cite{2015CQGra..32g4001T} and Virgo
\cite{TheVirgo:2014hva})  are sensitive to the relatively well understood signal from  the lowest-mass compact binaries
$M=m_1+m_2\le 16 M_\odot$ \cite{2003PhRvD..67j4025B,2004PhRvD..70j4003B,2004PhRvD..70f4028D,BCV:PTF,2005PhRvD..71b4039K,2005PhRvD..72h4027B,2006PhRvD..73l4012K,2007MNRAS.374..721T,2008PhRvD..78j4007H,gr-astro-eccentric-NR-2008,gw-astro-mergers-approximations-SpinningPNHigherHarmonics,gw-astro-PN-Comparison-AlessandraSathya2009}.  
Strong signals permit  high-precision constraints on binary parameters, particularly when the binary precesses.  
Precession arises only from spin-orbit misalignment; occurs on a distinctive timescale between the inspiral and orbit;
and produces distinctive polarization and phase modulations \cite{ACST,gw-astro-SpinAlignedLundgren-FragmentA-Theory,gwastro-SpinTaylorF2-2013}.  
As a result, the complicated gravitational wave signal from precessing binaries is unusually rich, allowing
high-precision constraints on multiple parameters, notably the (misaligned) spin
\cite{LIGO-CBC-S6-PE,gwastro-mergers-HeeSuk-FisherMatrixWithAmplitudeCorrections}.  
Measurements of the spin orientations alone could provide insight into processes that affect spin alignment, such as
supernova kicks \cite{2013MNRAS.434.1355J,2012MNRAS.423.1805N}, tides and post-Newtonian resonances
\cite{2013PhRvD..87j4028G}.  
More broadly, gravitational waves constrain the pre-merger orbital plane and total angular momentum direction, both of
which may correlate with the presence, beaming, and light curve \cite{2010ApJ...722..235V,2011ApJ...733L..37V,2013ApJ...767..141V} of any post-merger ultrarelativistic blastwave (e.g, short GRB)
 \cite{2009ARAA..47..567G}.  
Moreover, spin-orbit coupling strongly influences orbital decay and hence the overall gravitational wave phase: the
accuracy with which most other parameters can be determined is limited by knowledge of BH spins
\cite{1995PhRvD..52..848P,2013PhRvD..87b4035B,gwastro-mergers-HeeSuk-FisherMatrixWithAmplitudeCorrections,gwastro-mergers-HeeSuk-CompareToPE-Aligned}.  
Precession is known to break this degeneracy
\cite{2006PhRvD..74l2001L,2009PhRvD..80f4027K,2011PhRvD..84b2002L,gwastro-mergers-HeeSuk-FisherMatrixWithAmplitudeCorrections,2007CQGra..24..155V,LIGO-CBC-S6-PE}.  
In sum,  the rich  gravitational waves emitted from a precessing binary  allow higher-precision measurements of
individual neutron star masses, black hole masses,  and black hole spins, enabling  constraints on their distribution across multiple events.  In
conjunction with electromagnetic measurements, the complexity of a fully precessing gravitational wave signal may enable
correlated electromagnetic and gravitational wave measurements to much more tightly constrain the central engine of
short gamma ray bursts.

Accurately simulating richness and complexity comes at a price: essentially,\footnote{Recently, Kesden et al provided a
  new approach to evolving double-spin compact binaries, potentially enabling more rapid time- and frequency-domain
  solutions to the spin precesion and orbit equations.  Though possible to use this approach to generate waveforms in principle, no implementation is available in
  \texttt{lalsimulation} for
\texttt{lalinference} at the time of writing. } the additional computational weight of numerically
evolving several  ODEs for the spin and orbit dynamics.  
This cost places a substantial burden on attempts to reconstruct the sources of an observed wave, since data must be
systematically compared against all possible candidate signals  \citeMCMC{}.   
Owing both to the relatively large number of parameters needed to specify a precessing binary's orbit and to the
seemingly-complicated evolution,   Bayesian parameter estimation methods have only recently
\cite{gwastro-pe-systemframe,gw-astro-PE-lalinference-v1}  become efficient enough to  draw
inferences about gravitational waves from a statistically significant sample of generic precessing sources \cite{gwastro-pe-Tyson-AstroSample-MassGap2015}.  
At the time of writing, successful calculations of this type require run times on the order of days with existing computational resources. 

Except for a relatively small corner of parameter space, however, one black hole spin dominates any precession; the smaller mass
usually has a strongly suppressed impact on angular momentum evolution because black hole spin ($\mathbf{S}$) scales as
the black hole mass $m$ squared
($\mathbf{S} = m^2 \boldsymbol{\chi}$).  The dynamics of a single spin are well-understood \cite{ACST} and easily
approximated in the time and frequency domain.  In particular,  the
\STLong{F2} (\ST{F2}) \cite{gwastro-SpinTaylorF2-2013} approximation provides a  fast and accurate model for
binary inspiral, valid over a broad range of mass ratios and spins.  At the time of writing, parameter estimation using
\ST{F2} as a template requires roughly an order of magnitude less time on comparable resources, often of order minutes.  

Low-latency parameter estimation on these timescales enables transformative followup electromagnetic  observations
\cite{LIGO-2013-WhitePaper-CoordinatedEMObserving}.  
The most tantalizing proposed electromagnetic counterparts to compact binary merger are expected to be brief, potentially disappearing within
days if not much
sooner \cite{2012ApJ...746...48M,2014MNRAS.439..757G,2014MNRAS.437L...6K,2014MNRAS.437.1821M,2014ApJ...780...31T,2013ApJ...775...18B}.  
Given limited resources, reliable low-latency parameter estimation of gravitational wave signals will significantly enhance the science output of
multimessenger, time-domain astronomy.  
The  use of rapid single-spin templates like \ST{F2} may be a critical ingredient
in enabling these followup observations, if parameter estimation with this approximation provides suffciently robust
predictions.

In this paper, we systematically assess the accuracy of parameter estimation with \ST{F2}.    
The \ST{F2} approximation is computationally efficient because  (a) it uses only a single spin, eliminating three 
subdominant degrees of freedom and enabling fast, analytic solutions to the precession equations; and particularly (b)
because it constructs a highly efficient stationary-phase approximation to the
single-spin kinematics and gravitational wave emission.    Though these approximations introduce systematic errors by
neglecting higher-order post-Newtonian terms (e.g., associated with two-spin effects),
on theoretical grounds one expects these uncertainties to be comparable to the neglected post-Newtonian
terms.  In other words, on theoretical grounds one anticipates comparable differences between (i) the predictions
constructed using the standard adiabatic quasicircular inspiral models \STLong{T4} (\ST{T4}) and
\STLong{T2} (\ST{T2}) where both black holes are allowed to have generic spins \cite{BCV:PTF,2003PhRvD..67j4025B,gw-astro-PN-Comparison-AlessandraSathya2009}; 
(ii)  between \ST{T2} with two spins and \ST{T2} with only one nonzero spin, restricted to  the more massive onbject;
and 
(iii)  between  \ST{T2} with two or one spin  and \ST{F2}.  
As part of a larger study of parameter estimation on an astrophysically-selected sample of sources
\cite{gwastro-pe-Tyson-AstroSample-MassGap2015}, in this work we systematically evaluate these hypotheses and the
astrophysical impact that systematic biases introduce.   
This paper is organized as follows.
In Section \ref{sec:Sample} we introduce our parameter estimation study, describing the population of events used;
the specific models adopted to infer compact binary parameters; and the specific techniques we used to reconstruct each
posterior parameter distribution.
In Section \ref{sec:Biases} we introduce and employ standard statistical tools (i.e., Student's t disribution) to
identify and quantify systematic differences between the posterior parameter distributions arrived at by using different
waveform models.   
In Section \ref{sec:Astro} we demonstrate these systematic errors, though statistically significant, rarely
significantly impact our astrophysical conclusions, particularly because these errors are comparable to other systematic
uncertainties.   
In Section  \ref{sec:Conclude} we summarize our conclusions.  
An appendix \ref{ap:Stats} briefly reviews  statistical tools used in our analysis. 

\subsection*{Context and related work}

Our study is the first large-scale investigation of parameter estimation accuracy with approximate precessing
templates, using production-scale code and an astrophysically-motivated sample.  Several groups are pursuing complementary methods to accelerate parameter estimation for precessing binaries, including
both fast and accurate waveform models
\cite{2014PhRvD..90l4029K,gwastro-approx-precessing-YunesEtAl-NeutronStars-2013,gwastro-approx-precessing-YunesEtAl-Generic-2013,2015ApJ...798L..17C}
\cite{gwastro-mergers-IMRPhenomP}
\cite{gwastro-mergers-PE-ReducedOrder-2013,2013PhRvD..87l2002S,2013PhRvD..87d4008C},
and  alternative architectures for the likelihood function \cite{gwastro-PE-AlternativeArchitectures}. 

As an example of a fast but accurate  waveform model, Klein
and collaborators have developed methods to construct a highly faithful  SPA-like fourier transform of the line-of-sight
waveform $h(t,\hat{n})$ 
\cite{2014PhRvD..90l4029K,gwastro-approx-precessing-YunesEtAl-NeutronStars-2013,gwastro-approx-precessing-YunesEtAl-Generic-2013,2015ApJ...798L..17C}.
This accurate but technically sophisticated approach differs substantially from the simpler and more approximate
\ST{F2}, defined as a term-by-term stationary-phase-approximated fourier transform of $h(t,\hat{n})
=\sum_{lm}h_{lm}(t)\Y{-2}_{lm}(\hat{n})$, using a corotating-frame expansion of $h_{lm}(t)$ to avoid precession-induced
phase catastrophes.  
By design extremely faithful, the latest Klein et al. approximation \cite{2014PhRvD..90l4029K} has been shown to enable faster parameter estimation, allowing the
authors to assess hypotheses about precessing \emph{double}-spin binaries \cite{2015ApJ...798L..17C}.  
As another example, reduced-order-modeling and SVD methods in principle offer a robust and rapid procedure to efficiently approximate
\emph{any} waveform, reconstructing the signal from a sparse set of basis signals and (interpolated) functions of
parameters \cite{gwastro-mergers-PE-ReducedOrder-2013,2013PhRvD..87l2002S,2013PhRvD..87d4008C}.   
Given the challenge of high-dimensional interpolation, reduced order models have to date been  incrementally
applied to high-cost and high-precision but low-dimensional models like  nonprecessing EOBNR \cite{gw-astro-EOBNR-Calibrated-2009,gw-astro-EOBspin-Tarrachini2012}, an
inspiral-merger-ringdown waveform;   and IMRPhenomP \cite{gwastro-mergers-IMRPhenomP}, a cousin to \ST{F2} which includes
approximate merger and ringdown   (Raymond et al, private communication). 
At the time of writing, \ST{F2} is the fastest available waveform model including
precession.

After the cost of waveform generation, the computational cost of parameter estimation is dominated by the cost per
likelihood, which historically has been dominated by the cost of operating on and fourier transforming long arrays.   Recently, several methods have been proposed to perform this comparison more efficiently
\cite{gwastro-mergers-PE-ReducedOrder-2013,2013PhRvD..87l2002S,2013PhRvD..87d4008C,gw-astro-ReducedOrderQuadraturePE-TiglioEtAl2014,gwastro-PE-AlternativeArchitectures}, by interpolating some combination
of the the waveform or likelihood;  by adopting a sparse representation to reduce the computational cost of data
handling; or by organizing the calculation to maximize the reuse of intermediate results obtained from these long and costly
array operations.

\section{Parameter estimation on an astrophysical sample}
\label{sec:Sample}
Gravitational wave parameter estimation involves systematic comparison of  candidate waveforms to data \citeMCMC{}.
We use  the  \texttt{lalinference\_mcmc}
\cite{gw-astro-PE-lalinference-v1} parameter estimation code to infer posterior parameter distributions, a method and tool validated by
extensive prior studies  including
\cite{LIGO-CBC-S6-PE,gw-astro-PE-Raymond,gwastro-mergers-HeeSuk-CompareToPE-Aligned,gwastro-mergers-HeeSuk-CompareToPE-Precessing,2014PhRvD..89j3012W,2015PhRvD..91d3002L,gwastro-pe-Tyson-AstroSample-MassGap2015,2015ApJ...804..114B,2014ApJ...795..105S}.   
This code can construct synthetic data from and perform inference with any of a wide range of waveform modes present in
the \texttt{lalsimulation} software library.  In this study, we always generated our candidate precessing signals using
the \ST{T2} approximation described in \cite{2013PhRvD..88l4039N,2014PhRvD..89b4010H}. 
Using this fixed data set, we recovered parameters using \ST{T4}, described in
\cite{2003PhRvD..67j4025B,gw-astro-PN-Comparison-AlessandraSathya2009}, with
one or two spins; \ST{T2}, with one or two spins; and \ST{F2}. 
In all cases we used a leading-order (Newtonian) amplitude; 3.5 PN order in orbital phase; up to 3.5 PN in spin-orbit
and 2PN spin-spin terms; and treated all objects as point particles with black-hole
like couplings in, for example,  the quadrupole-monopole terms.  
This binary evolution is terminated prior to merger, either when it reaches the “minimum energy circular
orbit”; when the orbital frequency ceases to increase monotonically; or when the post-Newtonian $v/c$ is greater than unity.
While our simulated binaries start their evolution when twice their orbital frequency is  20 Hz, as in previous work we
specify binary spin parameters at a reference frequency $f_{\rm ref}=100\unit{Hz}$
\cite{gwastro-pe-systemframe,gwastro-mergers-HeeSuk-FisherMatrixWithAmplitudeCorrections,gwastro-mergers-HeeSuk-CompareToPE-Aligned,gwastro-mergers-HeeSuk-CompareToPE-Precessing}.  
In these analyses, we adopt a fiducial 3-detector network: advanced LIGO \cite{2015CQGra..32g4001T} and Virgo
\cite{TheVirgo:2014hva}, assumed to have analytic design-sensitivity gaussian noise
power spectra provided by \texttt{lalsimulation}; for example,  we adopted the advanced LIGO  zero-detuned high-power configuration \cite{LIGO-aLIGODesign-Sensitivity,2015CQGra..32g4001T}.  
To simplify our analysis,  following previous studies \cite{gwastro-mergers-HeeSuk-CompareToPE-Aligned,gwastro-mergers-HeeSuk-CompareToPE-Precessing,gwastro-pe-Tyson-AstroSample-MassGap2015} we also adopt a unique  preferred noise realization for all sources: exactly zero.
Data was always sampled at a rate significantly in excess of the Nyquist frequency, at a sampling rate that depended on the
source mass.

We apply these tools to a fixed set of 998 events, whose parameters has been selected as part of a large astrophysical
study first reported by Littenberg et al \cite{gwastro-pe-Tyson-AstroSample-MassGap2015}.   
The set's members were randomly selected, with masses $m_1,m_2$ uniform in a triangle with $m_{1,2}\ge 1M_\odot$ and
$m_1+m_2\le 30 M_\odot$; spin magnitudes $\chi_{1,2}$ uniformly distributed between $[0,1]$; and spin directions
independently and uniformly distributed on a sphere.   The set's members also have random positions  in the universe,
subject to the restriction that no source has network amplitude $\rho \le 5$ in two or more detectors.  
For each  member of the above set, we carried out parameter estimation with two spins using \ST{T2} and with one spin
via \ST{T4}.  To further validate our
results, each double-spin \ST{T2} analysis was performed twice.   In a followup investigation, we also carried out parameter estimation on a random subset of 250 events
using  the   one spin
\ST{T2} and \ST{T4} waveform models.    
Finally, since each Markov-Chain Monte Carlo instance returns between $900$ and $3000$  independent posterior samples, we standardize our
statistical treatment by randomly selecting 900 such samples from each run.

In this work, we will investigate these  results by comparing the estimated posterior distribution for each of the systems'
 parameters at 100 Hz \cite{gwastro-pe-systemframe}, focusing particularly on parameters in common to all models: the
 component masses $m_{1,2}$; the magnitude and orientation of the most significant spin; and the  angle $\theta_{JN}$ between the
 total angular momentum and the line of sight.  Models with only one significant spin lack parameters for the
 subdominant spin.

\subsection{Selection bias, spin priors, and their implications for parameter estimation}
We select an astrophysically-motivated population of injections, drawn from an
initially uniform distribution in volume and orientation and further constrained by a signal amplitude cut.  As expected with real searches for astrophysical sources,
this cut strongly favors distant, nearly-face-on sources ($\theta_{JN}\simeq 0-0.8$).  
For sources with small $\theta_{JN}$, the angle between $\vec{L}$ and the line of sight is therefore nearly constant
as the binary precesses.  As a result, as first described in \citet{gwastro-pe-Tyson-AstroSample-SelectionBias2015}, the \emph{detection-weighted}
astrophysical population therefore strongly favors sources which are barely if at all modulated by precession of
$\vec{L}$.   
As previous  studies have shown \cite{gwastro-pe-Tyson-AstroSample-MassGap2015,gwastro-mergers-PNLock-Distinguish-Daniele2015}, the spins and
masses of sources in these
nearly-umodulated configurations are much more difficult to constrain than sources which exhibit strong precession along
the line of sight \cite{LIGO-CBC-S6-PE,gwastro-mergers-PNLock-Distinguish-Daniele2015,gwastro-mergers-HeeSuk-CompareToPE-Precessing}.  

By construction, our astrophysical sample also includes relatively few binaries with two dynamically significant spins.   
Using ${\cal R}\equiv
|\mathbf{S}_1-\mathbf{S}_2|/|\mathbf{S}_1+\mathbf{S}_2|$ to measure the relative magnitude of the  spin difference,  of
the 2000 binaries in our sample, only 109 have  $1-{\cal R}<0.5$ and only 387 have  $1-{\cal R}<0.8$.  Therefore, by design
the dynamics of each binary in our synthetic population should be well-described by the dynamics of a single significant
spin -- an essential assumption of the \ST{F2} model.    
Under different but still plausible astrophysical assumptions, such as significant black hole birth spin and
preferentially comparable binary black hole masses, both spins will play a more significant role in typical binaries'
dynamics and gravitational wave signal.    

Previous studies  suggested higher harmonics had little impact on parameter estimation \cite{gwastro-mergers-HeeSuk-FisherMatrixWithAmplitudeCorrections,gwastro-mergers-HeeSuk-CompareToPE-Precessing}, once precession broke
degeneracies in the signal.  For this reason,  both our candidate signals and parameter estimation strategies use the leading-order
gravitational wave amplitude.  Due to the astrophysical selection process favoring  small misalignment angles between
the line of sight and $\vec{J}$, higher harmonics can play a relatively more important role in breaking degeneracies.
Not least because  \ST{F2} presently lacks higher harmonics,  we defer a detailed investigation of the
impact of higher harmonics to a subsequent study.

\section{Quantifying and understanding systematic errors}
\label{sec:Biases}

\subsection{Global indicators of systematic error}
\label{sec:SNRReduced}
\begin{figure*}
\includegraphics[width=\columnwidth]{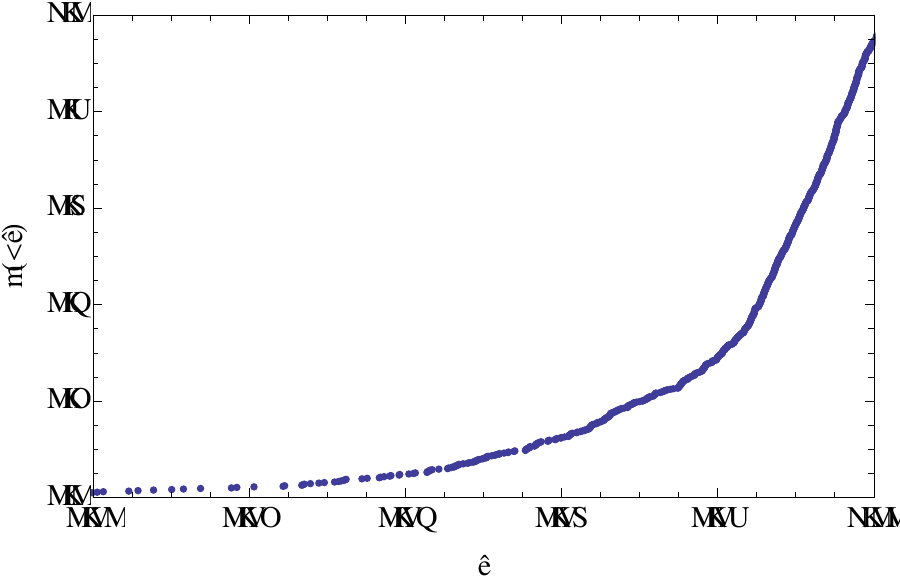}
\includegraphics[width=\columnwidth]{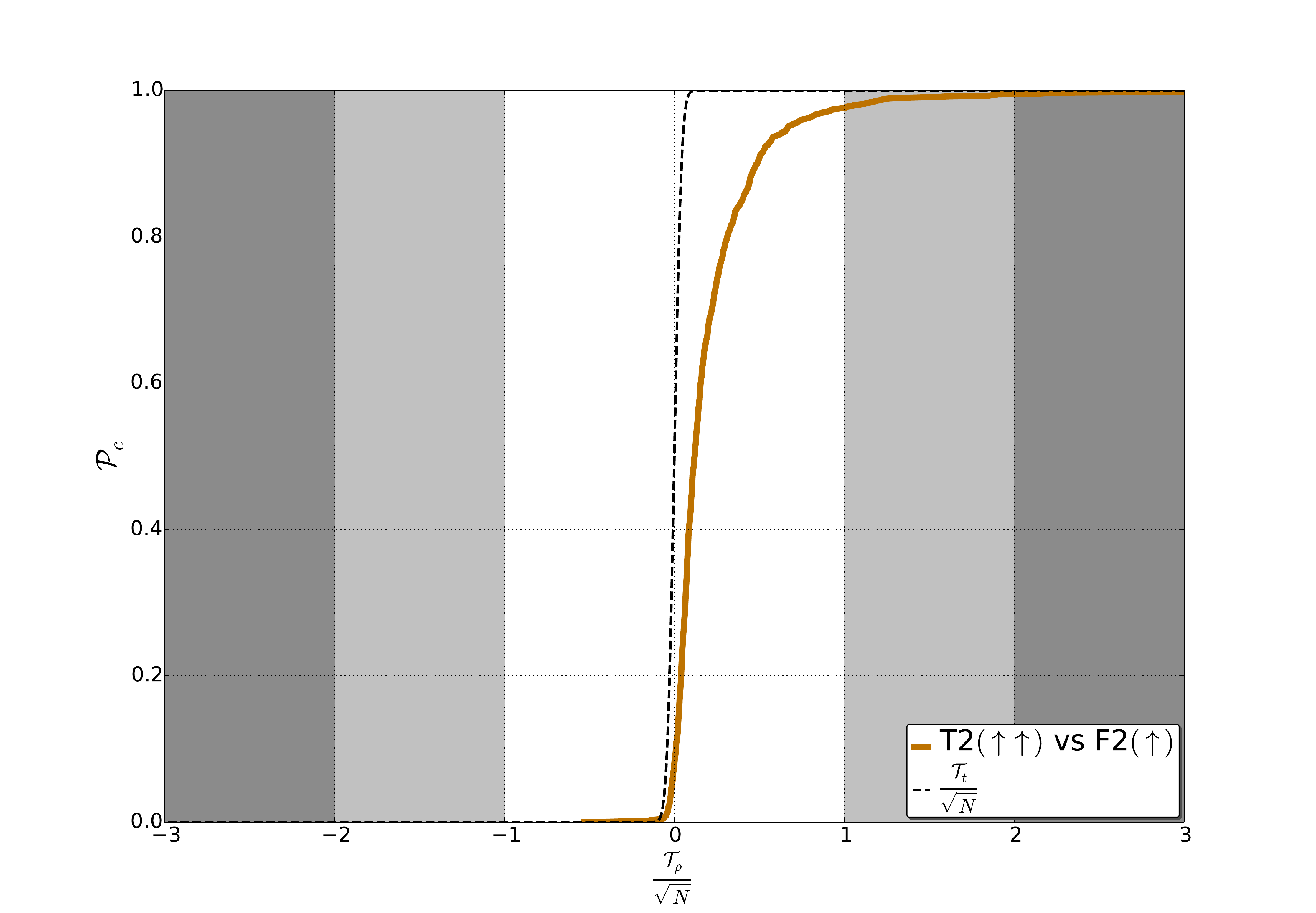}
\caption{\label{fig:SNRRatio}\textbf{SNR ratio distribution}: \emph{Left panel}: Ratio $r=(\bar{\rho}_{B})/(\bar{\rho}_A)$ of the largest SNR recovered
  for \texttt{STF2} to the SNR recovered for STT2, described in Sec. \ref{sec:SNRReduced}.  This distribution is significantly different from unity, with a
  magnitude consistent with the mismatch found in previous studies \cite{gwastro-SpinTaylorF2-2013}.
\emph{Right panel}:  Cumulative distribution of $\mathcal{T}_{\bar{\rho}}$ [Eq. (\ref{eq:TDefined})] divided by the square root
of $N$; in these units,
  the horizontal axis measures the
  difference in mean in units of the standard deviation.  The dashed curve is the theoretical cumulative $t$
  distribution.  This plot demonstrate that there is a
statistically significant difference between the average recovered SNR ($\rho$) by the two different models \ST{F2} and \ST{T2}.
}
\end{figure*}

The simplest way to demonstrate and quantify the existance of the systematic error is with the best-fit amplitude,
measured by the ``network amplitude''  $\rho \equiv \sqrt{2 \ln L}$, where $L$ is the likelihood (see, e.g., \cite{gw-astro-PE-lalinference-v1,gwastro-mergers-HeeSuk-CompareToPE-Aligned,gwastro-PE-AlternativeArchitectures}).    In gaussian noise, this quantity is nearly
normally distributed with unit mean in the presence of a signal, with mean ampitude proportional to the overlap
between the signal $h$ in the data and the best-fitting member of the signal manifold.     If two models $A$ and $B$ are applied to the same
data, known to contain a signal from $B$, and $\E{\cdot}$ denotes an expectation value, then the ratio
$\E{\rho_A}/\E{\rho_B}$ should be less than 1, reflecting ``mismatch'' between $A$ and $B$ associated with the inability
of the best-fitting members of $A$ to reproduce the signal in $B$.  For Markov-Chain Monte Carlo, this expectation can
be efficiently implemented as a direct sample average: if the MCMC has $N$ sample points with network amplitudes
$\rho_k$, then  $\bar{\rho}\equiv \frac{1}{N} \sum_k \rho_k$ is nearly equal to the expectation value.  

 Figure \ref{fig:SNRRatio} shows the results of this analysis, expressed as a cumulative distribution of $r\equiv
 \bar{\rho}_A/\bar{\rho}_{B}$, where $B$ is double-spin \ST{T2} and $A$ is \ST{F2}. 
This figure indicates significant differences between \ST{F2} and both single- and double-spin
\ST{F2}.  
In other words, for a significant fraction of sources, even the best fitting members of   \ST{F2} do not
completely  reproduce our zero-noise data.       Our results agree with the original investigations of \ST{F2}
\cite{gw-astro-SpinAlignedLundgren-FragmentA-Theory}.  
Though these mismatches allow systematic bias, not all systematic bias is astrophysically significant.  For example, Cho and
collaborators demonstrated that in many cases higher harmonics introduce a significant systematic bias in BH-NS
parameter estimation, almost
exclusively isolated to 
astrophysically irrelevant  parameters \cite{gwastro-mergers-HeeSuk-FisherMatrixWithAmplitudeCorrections,gwastro-mergers-HeeSuk-CompareToPE-Aligned,gwastro-mergers-HeeSuk-CompareToPE-Precessing}.

\subsection{Parameter biases}

\begin{figure}
\includegraphics[width=\columnwidth]{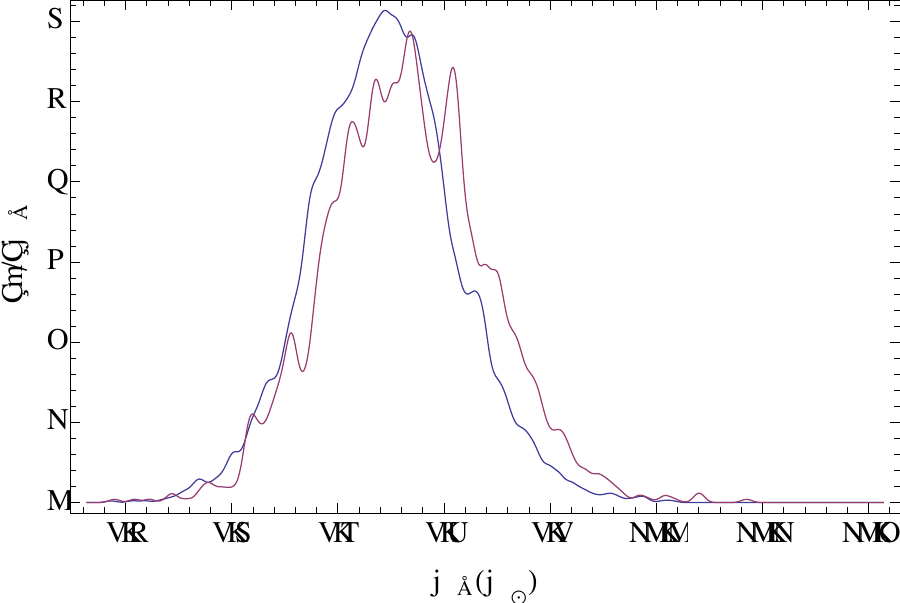}
\caption{\label{fig:PosteriorMchirpExample}\textbf{Example of posterior distributions}: Using one random example, this figure shows the one-dimensional distribution of chirp mass
  inferred from the \ST{F2} (red) and \ST{T2} (blue) distributions.  This figure shows that though the
  two methods give very similar results, the  two distributions are slightly
  offset.  Our study demonstrates these small offsets occur ubiquitously, at a statistically significant level. 
}
\end{figure}

As anticipated from the discussion above, the posterior distributions derived from \ST{F2} differ slightly
from the predictions produced using other approximations.  Figure \ref{fig:PosteriorMchirpExample} shows a
randomly-selected example.  In this figure and in general, the posterior distributions are qualitatively similar, except
for small systematic offsets typically  comparable to but smaller than the widths of each posterior distribution. 

To quantify the difference bewteen the two distributions' means, we use a tool from classical frequentist statistics:
Student's t.  As reviewed in the Appendix, for each simulation $k$ and each parameter $x$, we evaluate the sample mean
$\bar{x}_k=N^{-1}\sum_{\alpha=1}^N x_{\alpha,k}$ and sample standard deviation $s_{x,k}$, defined as $s_{x,k}^2 =
(N-1)^{-1}\sum_{\alpha} (x_{\alpha,k}-\bar{x}_k)^2$, where $N$ is the number of samples in each simulation and
$\alpha=1\ldots N$ indexes the posterior samples.   
(We require an equal number of samples from each posterior to simplify our interpretation; for a more general approach,
see the Appendix).   For each
pair of simulations of the same data $k$ with different
approximations $A$ and $B$, we then evaluate 
\begin{eqnarray}
\label{eq:TDefined}
{\cal T}_{x,k}(A,B) &\equiv& \frac{ \bar{x}_{k,A} - \bar{x}_{k,B}}{\sqrt{(s_{x,k,A}^2+ s_{x,k,B}^2)/N}}
\end{eqnarray}
Qualitatively speaking, the value of ${\cal T}_{x,k}(A,B)$ measures the difference in means between the two distributions,
scaled to the standard deviation divided by $\sqrt{\textbf{1000}}\simeq 30$; values less than or comparable to 30 are
therefore small compared to typical statistical errors. 
The distribution of ${\cal T}_{x}(A,B)$ should be nearly $t$-distributed with $2(N-1)$ degrees of freedom.  The empirical cumulative distribution of ${\cal T}_{x}(A,B)$ can  be evaluated by sorting the array of 984 ${\cal
  T}_{x,k}(A,B)$ values and compared, both to the theoretical $t$ distribution and to the results when the models $A$ and $B$ changes.

\begin{figure}
\includegraphics[width=1.1\columnwidth]{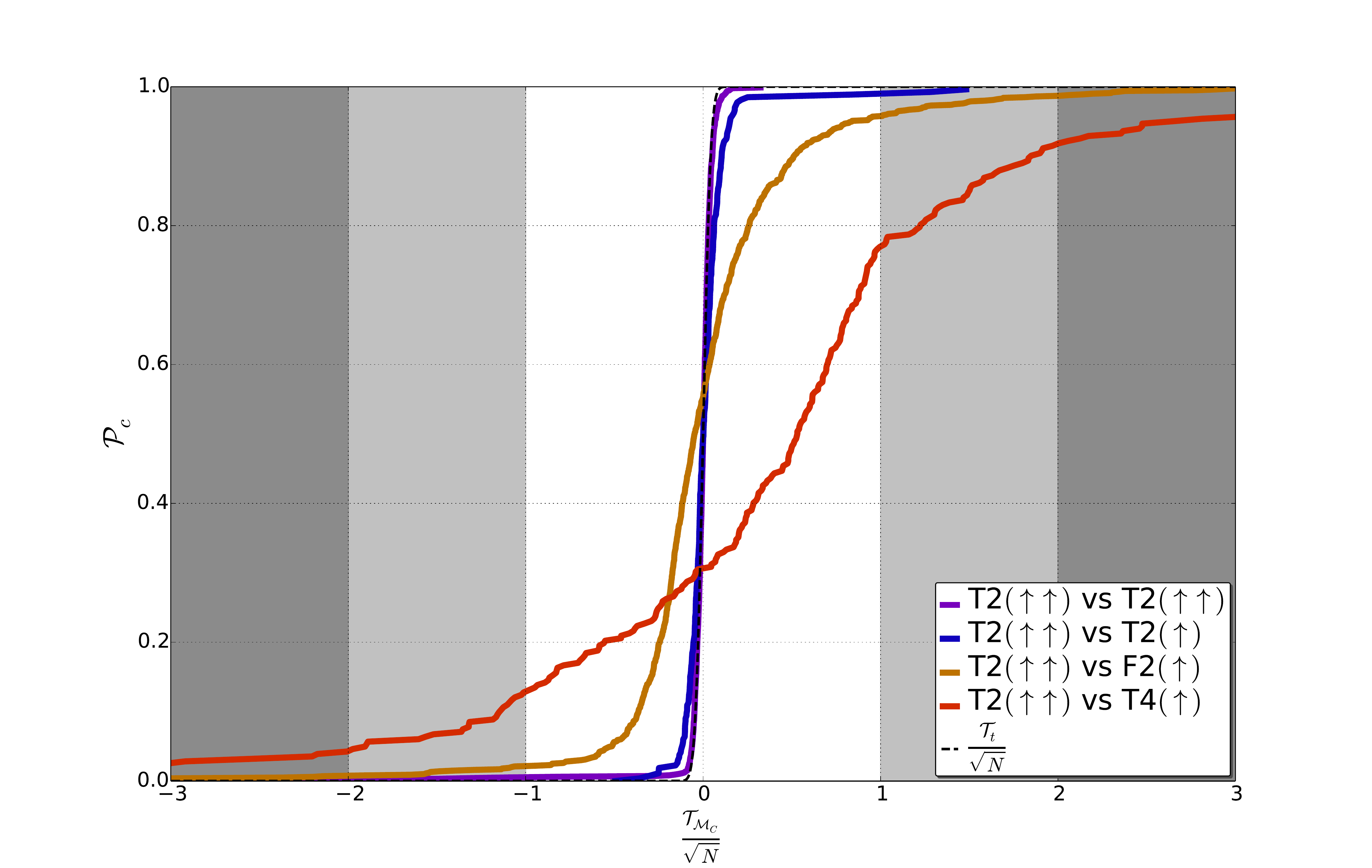}
\caption{\label{fig:TTest:Cumulative}\textbf{Scaled difference in mean chirp mass}: For the same data, we performed parameter estimation twice,
  reporting the cumulative distribution t-test scores in Eq. (\ref{eq:TDefined}) divided by $\sqrt{N}$.  In these units,
  the horizontal axis measures the
  difference in mean in units of the standard deviation.    The dashed curve shows the expected result, a $t$ distribution; the
  purple and blue curve shows the results when both models were \ST{T2} with the same or unequal numbers of nonzero spins;  the orange curve shows the results
  when comparing (double spin) \ST{T2} to \ST{F2}; and the red curve shows a comparison between single-spin \ST{T4} and
  double-spin \ST{T2}.
}
\end{figure}

Figure \ref{fig:TTest:Cumulative} shows the distribution of ${\cal T}_{\mc}$ for model $A$  being
\ST{F2},\ST{T2} double spin, \ST{T2} single spin with model $B$ being fixed to
\ST{T2} double spin.  
First and foremost, this model shows that when $A=B=$\ST{T2}, the distribution of ${\cal T}_{\mc}$ closely
follows the expected $t$ distribution.   
Second, the close agreement between a single and double-spin \ST{T2} model strongly suggests that a
single-spin model accurately reproduces most sources.   
Third and critically, this figure suggests that \ST{F2} differs substantially from both double- and even
single-spin \ST{T2}.  
These differences are smaller but still significant for low-mass and high-mass ratio  signals, where \ST{F2} by design
should be most accurate.

\begin{figure}[h!]
\includegraphics[width=\columnwidth]{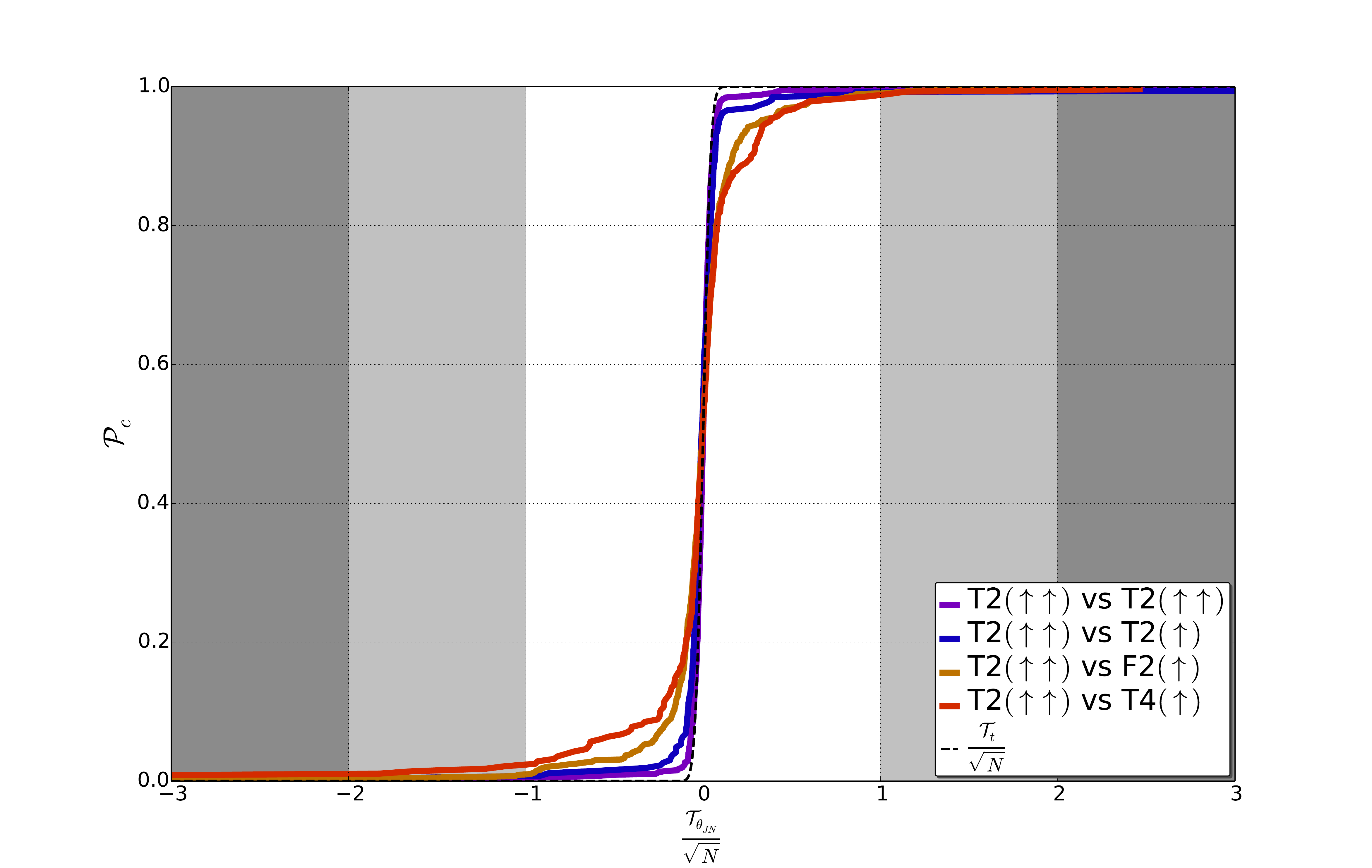}
\includegraphics[width=\columnwidth]{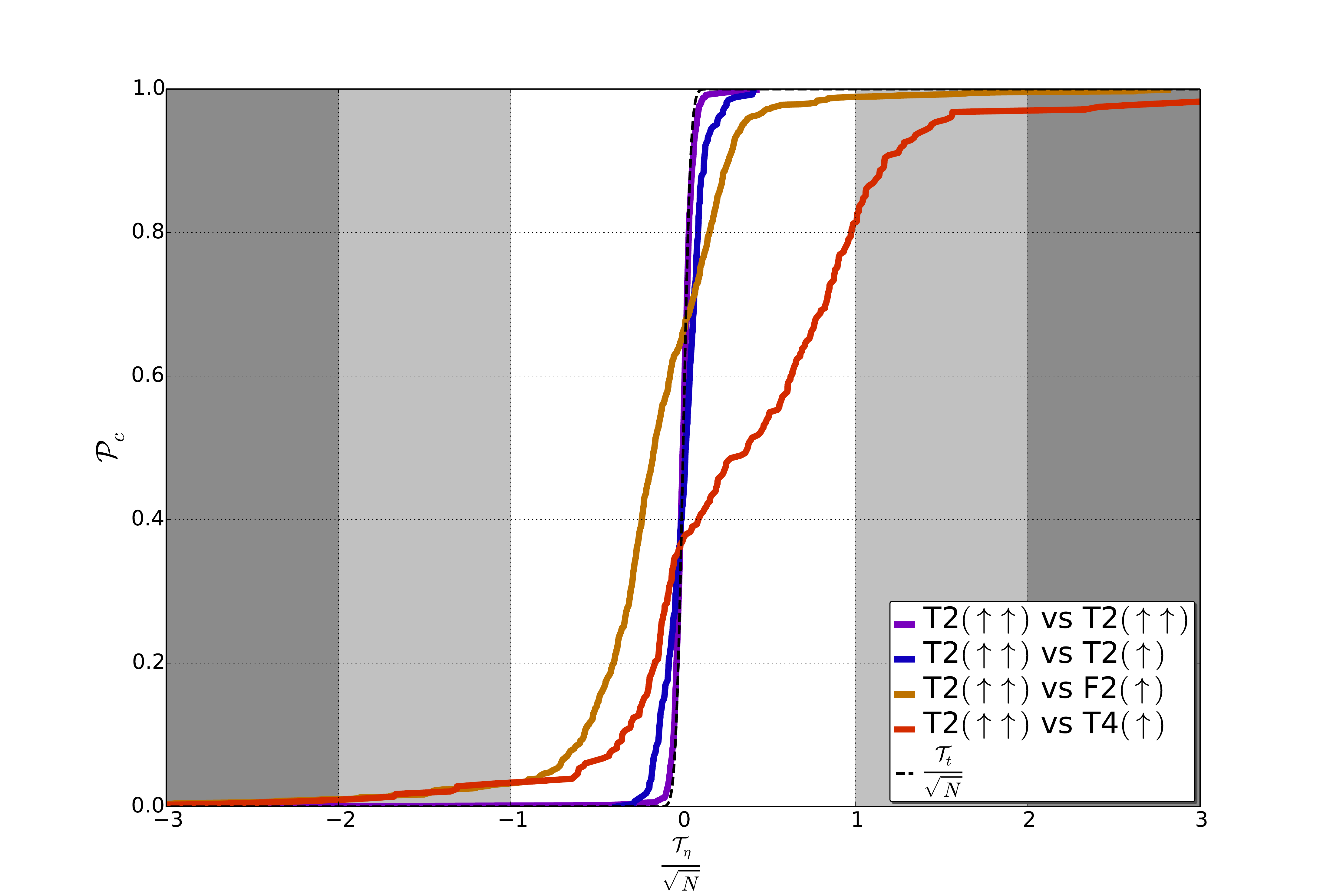}
\includegraphics[width=\columnwidth]{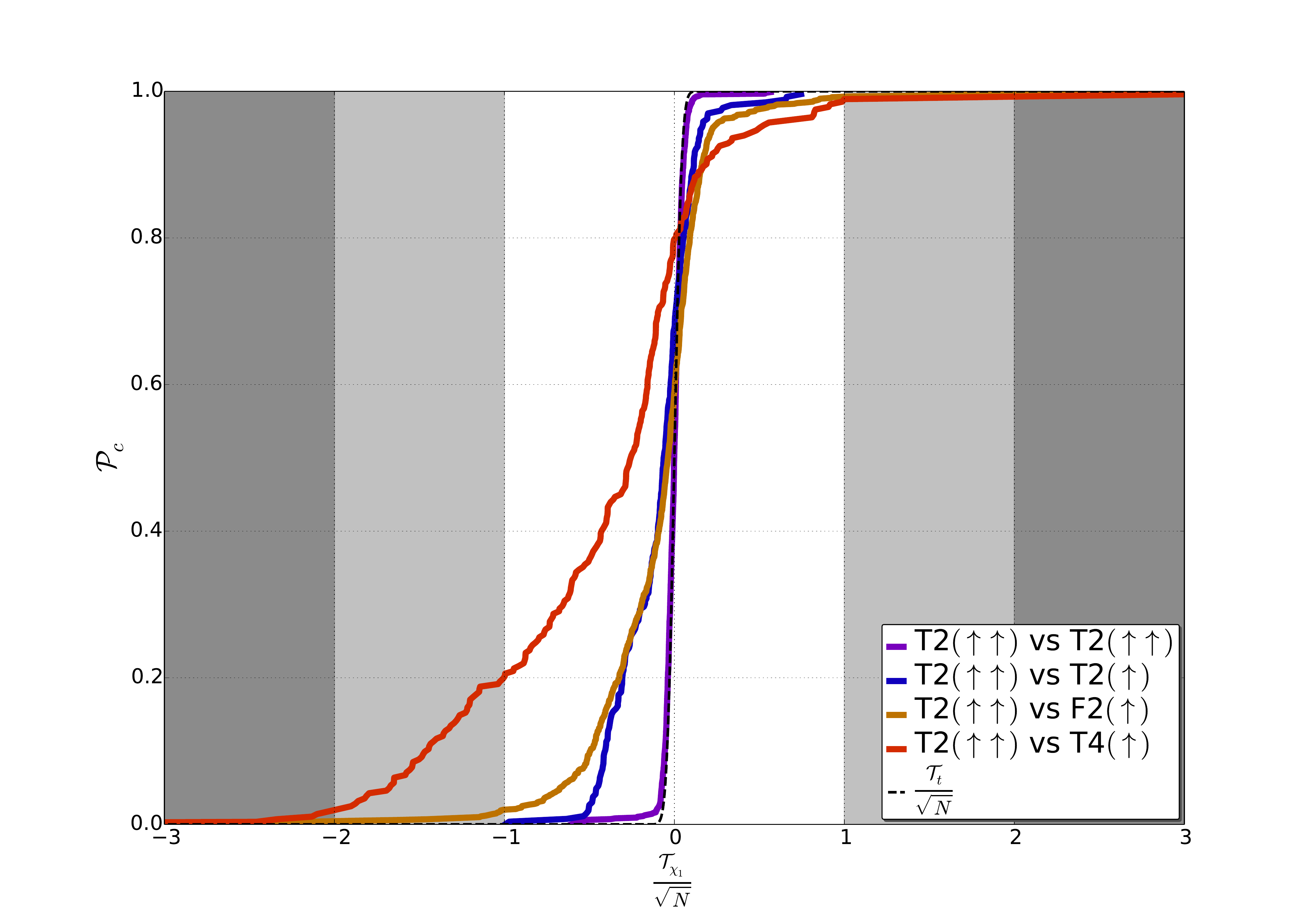}
\caption{\label{fig:PNSystematics:T4vsT2}\textbf{Results for additional parameters and PN approximants}: Like Figure \ref{fig:TTest:Cumulative}, a
  cumulative plot of ${\cal T}_x$ for $x=\chi_1,\eta,\theta_{JN}$.  The orange curve corresponds to comparisons between
  \ST{T4} (single spin) and \ST{T2} (double spin).
These orange curve  demonstrates that
  parameter estimation with post-Newtonian approximation schemes that adopt otherwise identical physics yield considerably different
  posterior distributions for \emph{intrinsic} parameters.  All approximations agree on geometric parameters like
  $\theta_{JN}$.
}
\end{figure}

Though small but statistically significant differences exist between \ST{F2} and time-domain approximations,
these differences are much smaller than the corresponding effect from  systematic uncertainty in the (orbital phase of the) post-Newtonian
approximation to precessing binaries.  
To illustrate the impact of systematic error in the post-Newtonian approximation, we perform the most conservative change possible: we
adopt identical physics and identical termination conditions, but construct our gravitational wave spin and orbit evolution using the \ST{T4}
scheme rather than the \ST{T2} scheme.  By design, these two methods must agree up to unknown higher-order
post-Newtonian (spin) terms in Taylor series for the factors of the (orbit-averaged) $dv/dt$ or $dt/dv$.    
Figure \ref{fig:PNSystematics:T4vsT2} shows our results, expressed  using the same strategy as described above to characterize differences between posterior means. 
 For this and all other parameters, the mean of the posterior
derived from \ST{T4}  is often farther from the mean of \ST{T2} than is the mean derived from \ST{F2}.   
Figure \ref{fig:PNSystematics:T4vsT2} provides a corresponding comparison using the intrinsic parameters $\eta,\chi_1$.  
As with the chirp mass, these distributions usually show significant systematic differences between
\ST{T2} and \ST{F2}.  Also like the chirp mass, these systematic differenes are  comparable to the systematic uncertainty seen between
\ST{T2} and \ST{T4}.  
This analysis suggests that the systematic errors introduced by restricting to \ST{F2} are relatively small, compared to the large systematic
error currently inherent in a post-newtonian approximation to typical merging binaries.

Despite often statistically significant differences between approaches, parameter estimation employing these distinct
approximations does produce consistent answers to questions
less sensitive to the precise orbital phase, including extrinsic parameters.  For example, Figure \ref{fig:PNSystematics:T4vsT2} also shows the models agree on $\theta_{JN}$, the relative orientation between $\vec{J}$ and the
line of sight.   The close agreement between these results can be understood on two grounds.  First, as has been described in the literature
\cite{gwastro-mergers-HeeSuk-FisherMatrixWithAmplitudeCorrections,gwastro-SpinTaylorF2-2013,2012PhRvD..86f4020B}, the angle
between $\vec{J}$ and the line of sight is closely related to the magnitude of precession-induced amplitude and phase
modulations induced by the precession of $\vec{L}$ around $\vec{J}$.  These modulations enter into the waveform as a
rotation, multiplicatively, and hence approximately decouple from the orbital phase due to seperation of scales.
Second and more broadly,  precession is a robust effect at leading order, treated identically in both
schemes.  
As discussed later, the ability to make robust statements about binary geometry just prior to merger may be critical in
identifying or ruling out candiate short GRBs for extensive EM followup, since the directions of strongest emission
should correlate with $\vec{J}$. %

\subsection{Extent of the confidence intervals}
As illustrated by example in Figure
\ref{fig:PosteriorMchirpExample}, different waveform approximations generate posteriors that differ in mean but agree in
shape, particularly width.   To
demonstrate this agreement quantitatively, we use another tool from classical frequentist statistics: the F statistic.
As reviewed in the Appendix, the $F$ statistic is a ratio of the sample standard deviations from two independent
experiments, to assess whether the two distributions have the same width.  For each pair of simulations of the same data
but different waveform models $A$ and $B$, and each parameter $x$, we evaluate
\begin{eqnarray}
\label{eq:F}
{\cal F}_{x,k}(A,B)  =  \frac{s_{x,k,A}^2}{s_{x,k,B}^2}
\end{eqnarray}
If both posterior distributions of $x$ are gaussian with the same mean, then ${\cal F}_{x}(A,B)$ should be
$F_{n_1,n_2}$ distributed with $n_1=n_2=(N-1)$ degrees of freedom in the numerator and denominator.  

\begin{figure}
\includegraphics[width=\columnwidth]{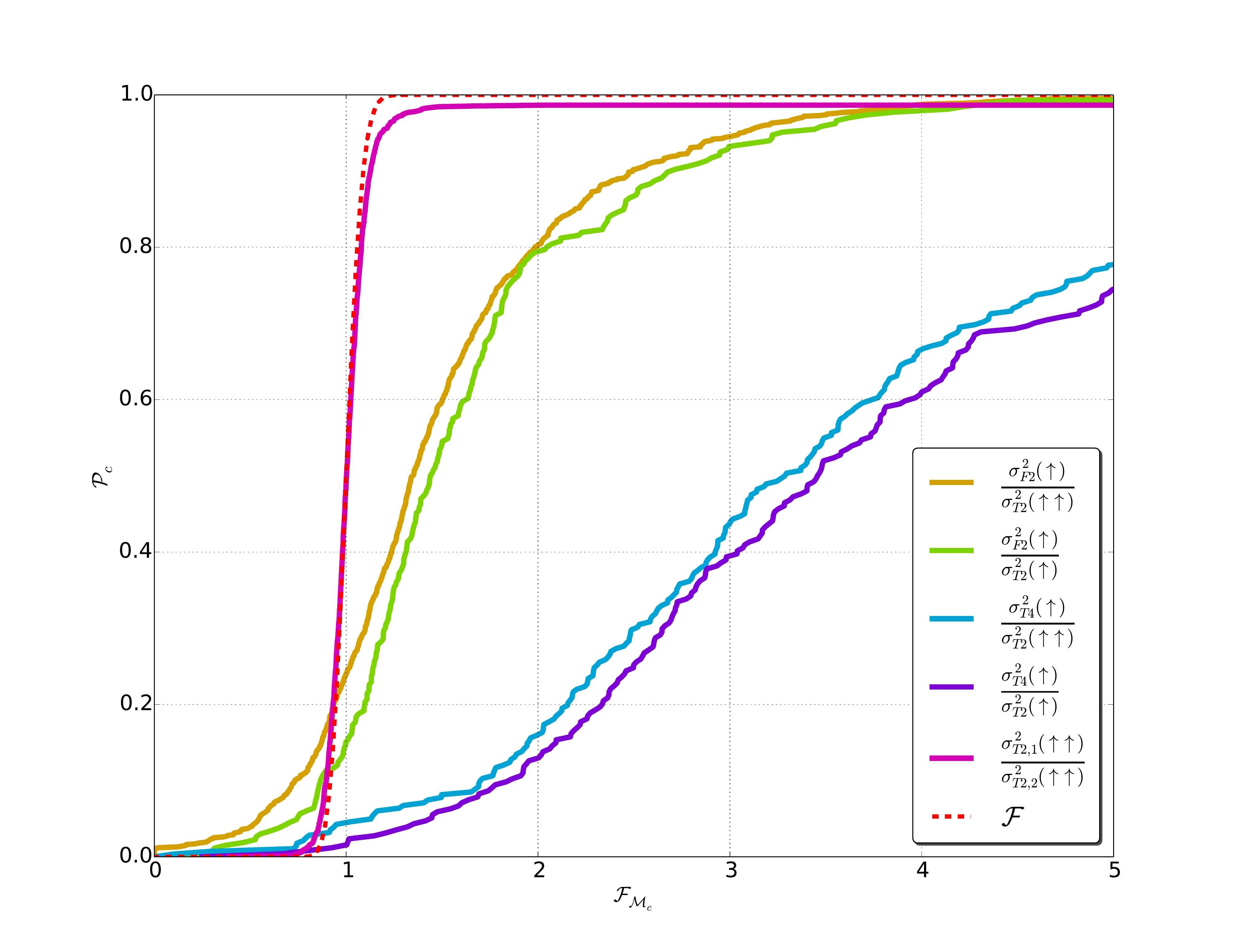}
\caption{\label{fig:FDistribution}\textbf{Simulation widths agree}: This plot shows that different approximations will
  predict posterior distributions of often significantly different widths.  The width of the distribution is strongly
  impacted by nongaussian tails and is a less-robust measure of distribution similarity. These results suggest that the
  posteriors produced by F2 are more similar to the ``true'' distributions than the results from T4. 
  [To emphasize the differences in distribution widths, we intentionally selected the parameter with the most gaussian
  distribution: the chirp mass. This test is less well suited to posterior distributions with significant nongaussian
  tails and cutoffs, notably  $\eta$ and $\chi_{1,2}$.]
}
\end{figure}

Figure \ref{fig:FDistribution} shows the empirical distribution of ${\cal F}_\mc(A,B)$ for $B=$ \ST{T2}
and $A$ being
\ST{F2},\ST{T2} double spin, \ST{T2} single spin with model $B$ being fixed to
\ST{T2} double spin.     These results suggest that adopting a different approximation often yields a posterior distribution
with a slightly different width.  These differences can be ascribed in part to the different best-fitting point: the
Fisher matrix and posterior distribution varies significantly across the parameter space.  These differences can also be
ascribed to the strong nongaussianity inherent in these distributions: the second moment used in the ${\cal F}$
distribution is sensitive to rare outliers.   For these reasons, a detailed study of the
change in extent and shape of the posterior distribution is substantially beyond the scope of this work.  For the
purposes of this study, we highlight the value of  \ST{F2} posterior distributions by two facts.  First, that the posterior distribution obtained with either a single
spin or \ST{F2} typically differs in width by a few tens of percent from the full posterior distribution, comparable to
the typical statistical error associated with the signal amplitude and only slightly greater than the sampling error
($\sqrt{1000}$) associated with our finite MCMC sample.  Second,
posterior distributions derived when using \ST{T4} as a template are significantly broader than posterior distributions
derived by any other means.   In short, while the posterior distribution with \ST{F2} is wider, we again assert that
differences associated with using \ST{F2} are larger than other systematic errors in our problem.

\subsection{Unique evidence for two spins?}

Lacking all degrees of freedom, a single-spin model like \ST{F2} cannot reproduce precession-induced modulations induced by the
subdominant spin.   Our sample of events strongly favors face-on sources ($|\cos \theta_{JN}|\simeq 1$), with minimal
modulation from the \emph{secular} precession of $\vec{L}$ around $\vec{J}$.  That said, our sample includes a
significant fraction of
comparable-mass BH-BH binaries with large and misaligned spins.  The relative precession of the three angular momenta
introduces additional modulations to  $\vec{L}$ \cite{ACST,2015PhRvL.114h1103K} and hence to $h(t)$,  potentially
communicating more observationally-accessible information about the spins, including the subdominant spin.  

\begin{figure*}
\includegraphics[width=\textwidth]{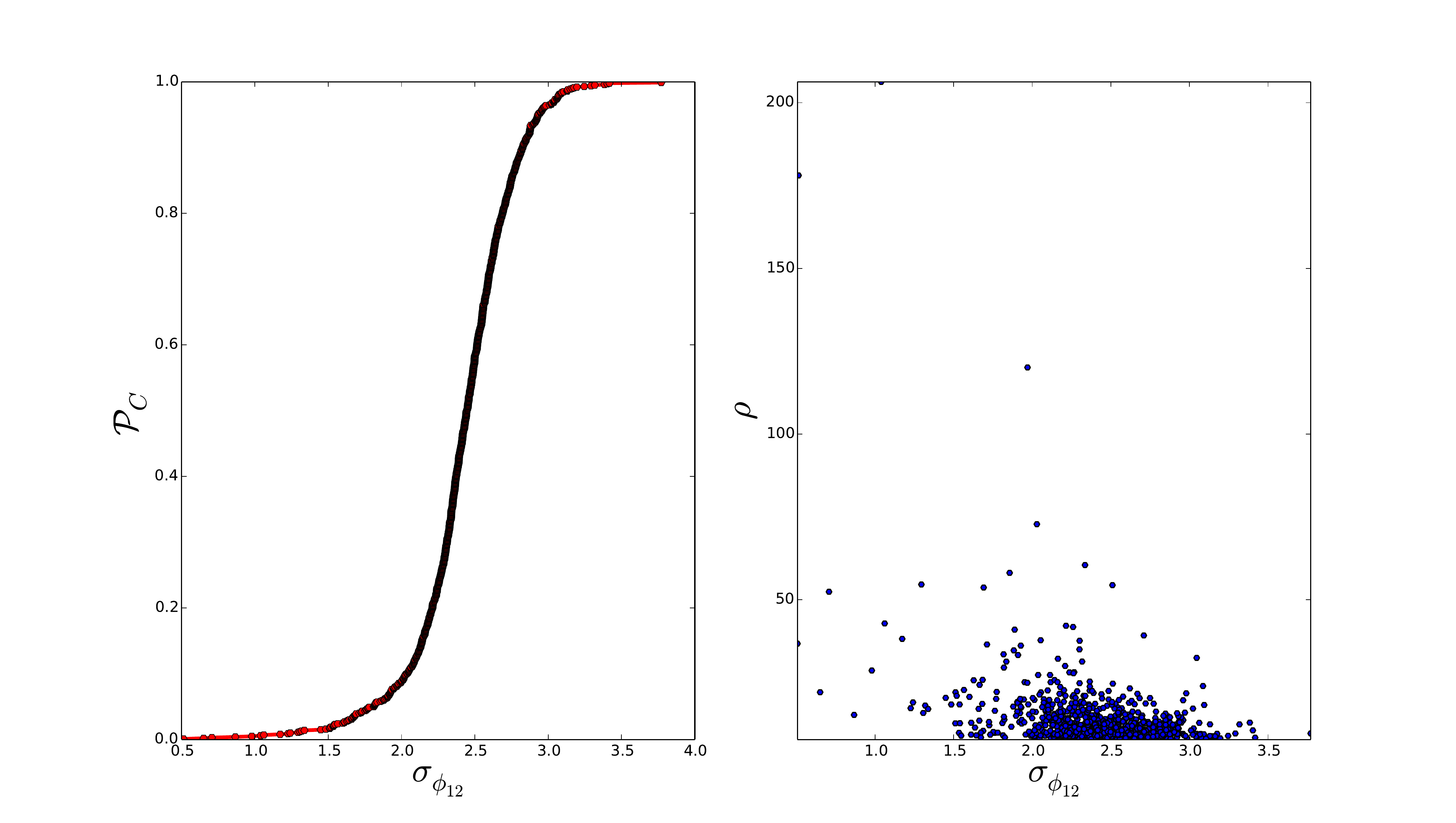}
\caption{\label{fig:TwoSpinPhi12} \textbf{Distribution of $\phi_{12}$}: Report on parmeter measurement accuracy of
  $\phi_{12}$ using two-spin \ST{T2}. [The parameter $\phi_{12}$ does not exist in an single-spin model, including  \ST{F2}.] \emph{Left panel}: Cumulative distribution of
  $\sigma_{\phi_{12}}$ [Eq. (\ref{eq:Phi12:Dev})].  For most cases, $\phi_{12}$ is approximately uniformly distributed.
\emph{Right panel}: Scatter plot of $\sigma_{\phi_{12}}$ versus $\rho$.  Sources with high amplitude provide more information about all parameters, including $\phi_{12}$. In some
exceptional cases -- with high amplitude, large spin, and fortuitious orientation  --  $\phi_{12}$ can thereby be
tightly constrained.
}
\end{figure*}

The close agreement between single- and double-spin \ST{T2} discussed above [Figure \ref{fig:SNRRatio}]  strongly suggests that the subdominant spin
rarely communicates observationally accessible information.  To quantitatively and directly assess whether the
subdominant spin's relative orientation can be measured,   we examine the distribution of
$\phi_{12}$, the angle between $\vec{S}_1$ and $\vec{S}_2$ in the plane perpendicular to $\vec{L}$
\cite{gwastro-pe-systemframe}.    As expected, in most cases  the posterior distribution of $\phi_{12}$ is nearly
uniform; however, a significant fraction of events ($\simeq 10\%$) have a nonuniform posterior distribution.  These concentrated posterior distributions are associated with larger-than-average signal amplitude. To measure this effect quantitatively without adopting a preferred range of the periodic
variable $\phi_{12}$, we calculate the following quantity: 
\begin{subequations}
\label{eq:Phi12:Dev}
\begin{eqnarray}
z\equiv \sum_\alpha e^{i \phi_{12,\alpha}} \\
\sigma_{\phi_{12}}^2 \equiv \sqrt{-2\ln |z|}
\end{eqnarray}
\end{subequations}
In the limit that $\phi_{12}$ is narrowly distributed near some preferred value, $\phi_{12}$ is the standard deviation
of the posterior.  
Figure \ref{fig:TwoSpinPhi12} shows the distribution of $\sigma_{\phi_{12}}$.   Based on  human followup, the handful of
cases with $\sigma_{\phi_{12}}\lesssim 1.8$ have
posteriors that slightly or significantly favor some range of  relative angles: the impact of the subdominant spin is
measurable. 

Not all large signal amplitudes are associated with binaries with two dynamically significant spins, nor with lines of
sight that facilitate the measurement of both spins.  Nonetheless, as one would expect, binaries with narrow posterior
distributions of $\phi_{12}$ are associated with unusually large signal amplitudes. 

\section{Astrophysical implications of systematic errors in low-latency followup with \texttt{STF2}}
\label{sec:Astro}

Having demonstrated the rapid but approximate parameter estimation enabled by \ST{F2} introduces small but
measurable systematic errors, we assess the practical astrophysical impact these errors introduce.   Low-latency
parameter estimation for precessing binaries facilitates electromagnetic followup by answering three critical questions
about BH-NS binaries.  First,  is the smaller object a neutron star \cite{2013ApJ...766L..14H,2015MNRAS.450L..85M,2014PhRvD..89j2005O,gwastro-pe-Tyson-AstroSample-MassGap2015}?  Second,  is the
total angular momentum pointing towards the observer?  Third and finally,  are the BH and NS masses and spins consistent with tidal
disruption prior to merger?  
We have used our large sample to evaluate the impact of systematic error on these three critical questions for
low-latency parameter estimation.

\subsection{Neutron star present}
\begin{figure}
\includegraphics[width=\columnwidth]{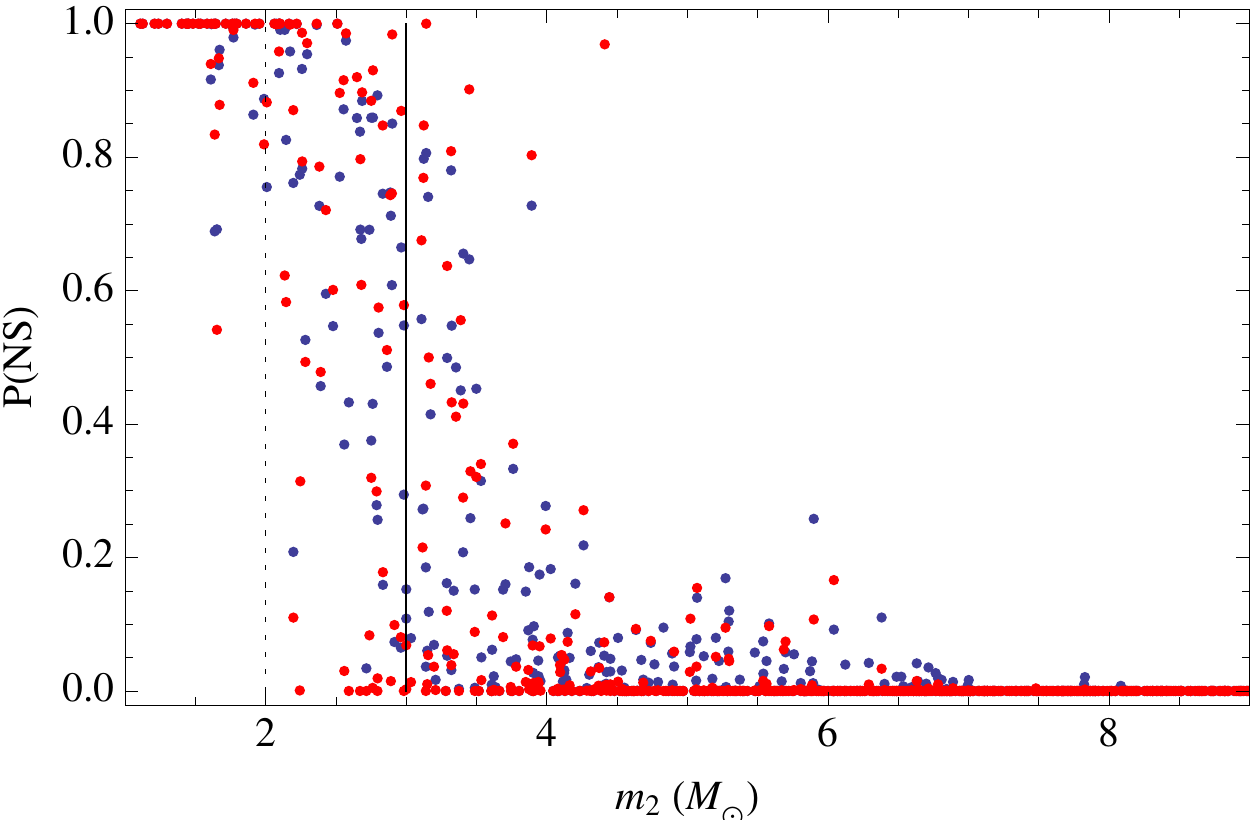}
\caption{\label{fig:AstroImplications:NSIdentify}\textbf{Identifying a candiate NS companion}: %
 Plot of the posterior probability that the secondary is a neutron
  star versus the smaller object's mass, as estimated using STT2 ($P_{A}$; blue) and STF2($P_B$; red).  
}
\end{figure}

For the purposes of discussion, we will call a smaller object a ``neutron star candidate'' if its mass is less than $3$
solar masses.  %
Using the estimated posteriors derived from both models $A$ and $B$
(STT2 and STF2, respectively) and this condition, we evaluate  posterior probabilities $P_A(NS)$ and $P_B(NS)$ that this
condition is satisfied, for all simulations.   
As seen in Figure \ref{fig:AstroImplications:NSIdentify}, the two probabilities largely agree, particularly when the NS mass is below $2 M_\odot$.
Given these results and systematic uncertainties in post-Newtonian waveforms with spin, electromagnetic followup will
likely occur for all sources with $P(NS)>0.1$.  In this scenario, our calculations suggest that roughly 65-70\% of all
followed-up sources will actually be BH-NS binaries; the same result holds for either model.

\subsection{Total angular momentum direction}
Both based on electromagnetic (jet break) and event rate arguments, short gamma ray bursts are assumed to be tightly
beamed into a solid angle $\theta\lesssim 20^\circ$ \cite{2014ARAA..52...43B}.    For a precessing binary, we will
conservatively assume the radiated energy is beamed along the total angular momentum direction -- for example, because
it is powered by accretion of matter onto the final black hole, whose total angular momentum direction is nearly
identical to the nearly-conserved total angular momentum direction of the binary from which it formed.  
Low-latency parameter estimation of gravitational waves can estimate the degree of misalignment $\theta_{JN}$ between
the line of sight and the total angular momentum direction of the progenitor binary.  

\begin{figure}
\includegraphics[width=\columnwidth]{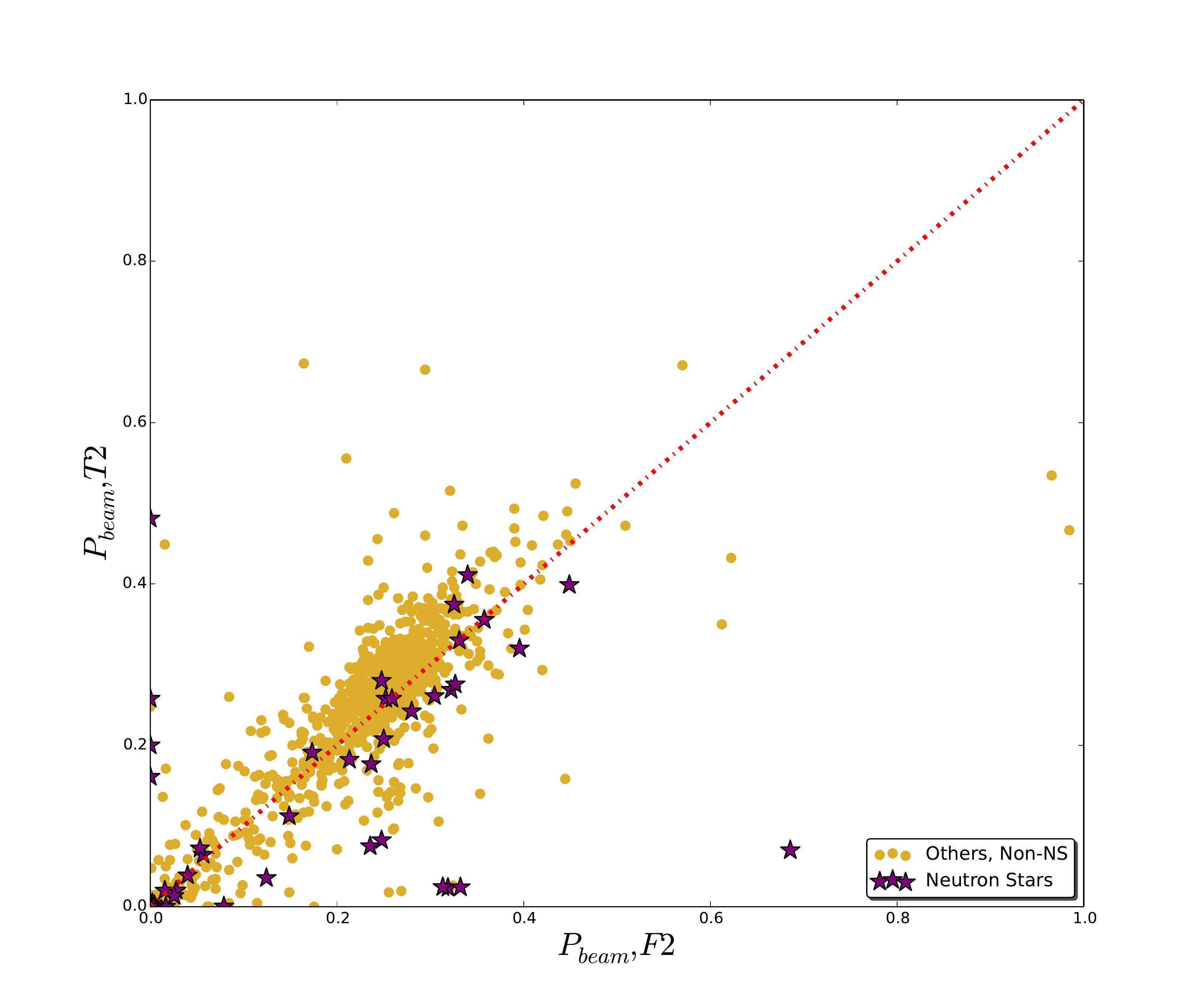}
\caption{\label{fig:AstroImplications:BeamingProbability}\textbf{Aligned with the line of sight}:
  Distribution $P_A(beam)$ (red) and $P_B(beam)$ for the probability of alignment between the line of sight
  and the total angular momentum direction, shown both for all elements of the astrophysical sample (yellow) and for
  the sources with neutron star compansions (purple).
}
\end{figure}

For the purposes of discussion, we will call a binary ``aligned with the line of sight'' if $\theta_{JN}<20^\circ$ or
$\theta_{JN}>\pi-20^{\circ}$.  This choice is highly arbitrary: neither theory nor observations motivate any hard
cutoff, though all disfavor extreme misalignment.  Using the estimated posteriors derived from both models, we define a probability $P_A(\text{beamed})$
and $P_B(\text{beamed})$ for each event that the binary is pointed towards us.     Figure
\ref{fig:AstroImplications:BeamingProbability} shows the distribution of these probabilities for all
events and for binaries containing a neutron star. %
Particularly given astrophysical
systematic uncertainty in the choice of cutoff angle, these distributions strongly suggest the beaming probabilities derived
from our two A and B models are nearly equivalent.   
The handful of cases with neutron star companions with inconsistent beaming probabilities ($P_{\rm beam}(F2)\simeq 0$ but
$P_{\rm beam}(T2)>0.1$ or vice-versa) were associated with
either   highly asymmetric  ($m_1\gtrsim  20 M_\odot$) or  nearly edge-on binaries.

\subsection{Tidal disrupt prior to merger}
To provide an unambiguous albeit approximate quantity to identify candidate tidal disruption events in binary mergers, we employ the
following procecure to identify posterior samples consistent with tidal disruption of a neutron star.  First, the
smaller mass must lie within the range of masses allowed by our fiducial NS equation of state,
 here
between $0.5 M_\odot$ and $2.5 M_\odot$.  Second, the disruption process must leave behind a remnant disk with nonzero
mass, as estimated using Foucart's expression \cite{2012PhRvD..86l4007F} (his Eqs. (6,12-13)):
\begin{align}
\frac{M_{disk}}{M_{NS}} &=
 0.288 (3 m_{bh}/m_{ns})^{1/3}[1-2 {\cal C}] 
\nonumber \\
&
 -0.148 (m_{bh}/m_{ns})  {\cal C} R_{isco}(a)/m_{bh} 
\label{eq:Mdisk}
\end{align}
where $q=m_{bh}/m_{NS}$ is the binary mass ratio; $a=S_{bh}/m_{bh}^2$ is a dimensionless measure of the black hole spin;
and  where ${\cal C}(m_{ns})=m_{ns}/R_{ns}$ is a mass- and equation-of-state dependent  measure of the neutron star
compactness.  In Foucart's expression, shown above, $R_{isco}(a)$ is the radius of the  innermost stable equitorial circular orbit  of a test particle
about a Kerr black hole  \cite{1972ApJ...178..347B}. %
When the black hole spin is not aligned with the orbit, we use the black hole spin magnitude $a$ in this
expression.\footnote{We adopted an orientation-independent expression for tidal disruption probability for simplicity
  and to maximize the number of binaries in our sample which satisfy this condition.  While physically more appropriate choice for the black hole spin would  $a=\hat{L}\cdot
  S_{bh}/m_{bh}^2$ in this expression, extremely few 
}
Using the fraction of all samples which satisfy this condition, we arrive at an (equation-of-state-dependent)
probability that the candidate event produces a tidal disruption.  Our Bayesian  approach  generalizes previously-reported
  Fisher-matrix-based  \cite{2014PhRvD..89f4056M} or search-template-based
  \cite{2014ApJ...791L...7P} estimates, incorporating state-of-the-art posterior estimates of each compact binary's
  parameters.  
Using \ST{F2}, these estimates can be evaluated in  one to a few hours using current sampling algorithms in
\texttt{lalinference\_mcmc}, with further performance improvements expected before the first few detections.

\begin{figure}
\centering
\includegraphics[width=\columnwidth]{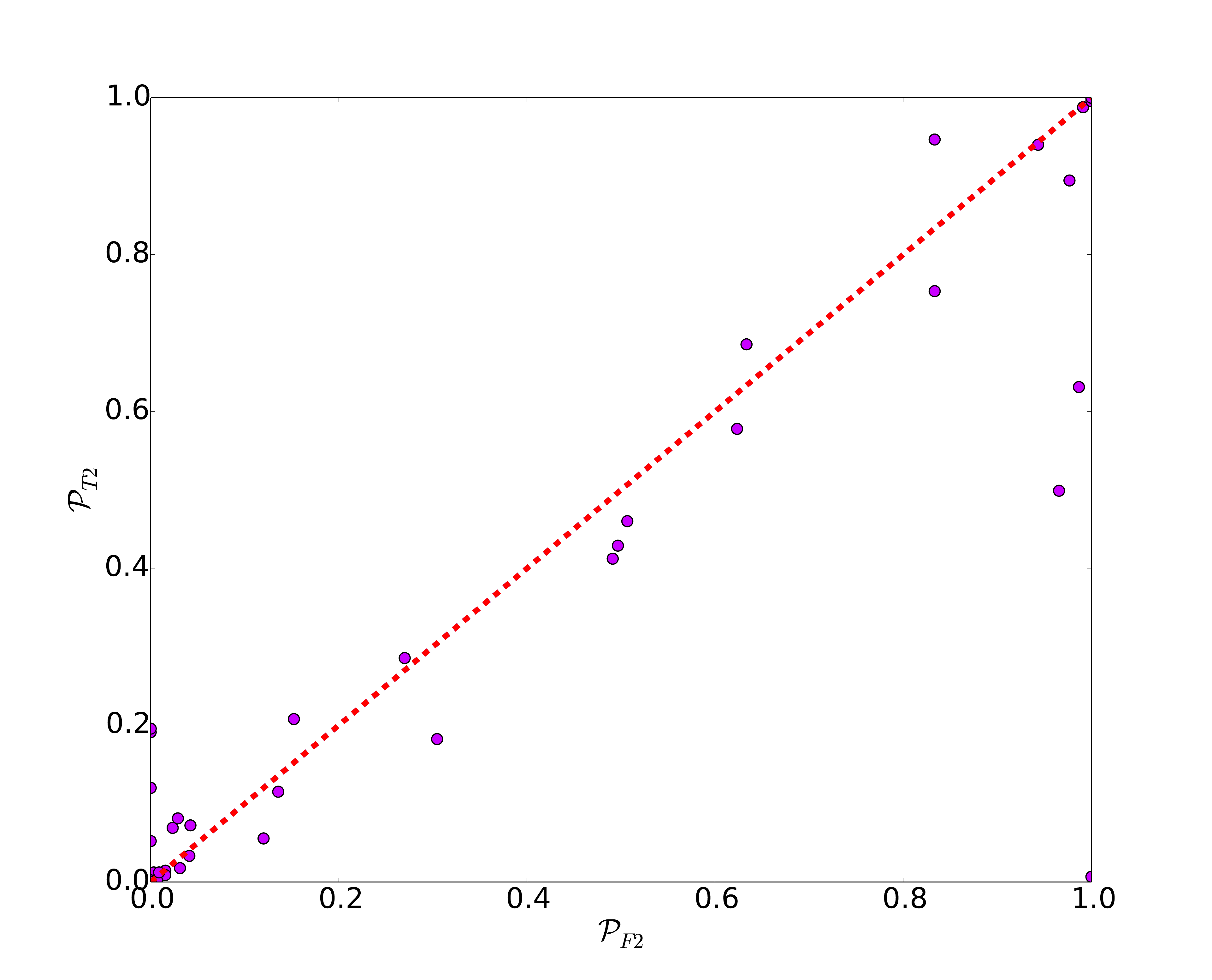}
\caption{\label{fig:TidalDisrupt}\textbf{Tidal disruption probability}: Plot of the posterior tidal disruption probability evaluated via
  Eq. (\ref{eq:Mdisk}) using parameters drawn from \texttt{SpinTaylorF2} [$P_{F2}$] and using parameters drawn from
  \texttt{SpinTaylorT2} [$P_{T2}$].}
\end{figure}

Figure \ref{fig:TidalDisrupt} shows a scatterplot of the tidal disruption probabilities evaluated using the posterior
parameter distribution derived using \ST{F2} and \ST{T2}.  As expected from Figure
\ref{fig:AstroImplications:NSIdentify}, the two approaches do not report precisely the same posterior probabilities for
each event.  Given systematic uncertainty in the nuclear equation of state and in post-Newtonian models for precessing
binaries, however, the two models report adequately  similar probabilities for targeted electromagnetic
followup.  %

\section{Conclusions}
\label{sec:Conclude}

Motivated by the need for low-latency parameter estimation, in this work we estimate the systematic error  introduced
into posterior parameter distributions and astrophysical predictions  by employing a rapid but approximate \ST{F2} waveform
model in favor of time-domain  models like \ST{T2} which include more physics.  Though statistically significant
differences exist, we demonstrate by repeated examples that these differences are small compared to the systematic
post-Newtonian modeling uncertainty inherent in our present, still-approximate understanding of compact binary
inspiral.  
Because the \ST{F2} model includes only one dynamically significant spin,  our results also consistent with suggestions that only properties
associated with a single effective spin will be observationally accessible with the first few gravitational wave
detections.   
Finally, we show that while different approximants disagree on \emph{intrinsic} parameters, the approximants lagely
agree on extrinsic, geometric parameters like $\theta_{JN}$.  The stability of extrinsic parameter estimates will be
important for targeting limited electromagnetic followup resources.

Our compelling results are in no way in conflict with prior and concurrent  studies which demonstrated that, in specific
moderate-amplitude cases, both black
hole spins can be independently constrained in mangitude and direction.  For the detection-weighted
astrophysical sample adopted in these studies, most sources have only one significant spin (e.g., due the mass and spin
prior) and have total angular momenta nearly along the line of sight.   As described elsewhere \cite{gwastro-pe-Tyson-AstroSample-SelectionBias2015},
these circumstances minimize the ability of precession-induced-modulations to break degeneracies and enable
both spins to be measured.  
Further investigations would be needed to determine if a different prior, favoring comparable-mass high-spin black hole
binaries, will enable high-precision measurement of both black hole parameters.

Our analysis employed inspiral-only waveforms which lack the coalescence and ringdown signals present in real binary
black hole merger signals.  Their unphysical termination conditions are known to introduce
convention-dependent artifacts into  parameter estimation, with increasing impact as the total binary mass increases
\cite{2014CQGra..31w5009C,2014CQGra..31o5005M}.   
A detailed discussion of waveform termination conditions is beyond the scope of this paper.  That said, we anticipate
that waveform termination conditions do not dominate the differences we observe.  We found similar
cumulative distributions when examining only low-mass sources, for which the impact of termination conditions is
reduced. 

\noindent \emph{Acknowledgements}:   TBL  acknowledges NSF award PHY-
1307020.  BF was supported by the Enrico Fermi Institute
at the University of Chicago as a McCormick Fellow. Computational
resources were provided by the Northwestern University
Grail cluster through NSF MRI award PHY-1126812.

\appendix
\section{Review of statistics}
\label{ap:Stats}

The Student-t distribution arises from the distribution of the  ratio of $t=x/\sqrt{z/n}$ where $x$ is normally
distributed with zero mean and unit variance and where  $z$ is an
  independent $\chi^2$-distributed variable with $n$ degrees of freedom.  After some algebra, the probability
  distribution function (PDF) of $t$ is 
 \begin{align}
p(t) &= \frac{\Gamma(\frac{n+1}{2})}{\sqrt{\pi n}\Gamma(n/2)} \frac{1}{(1+\frac{t^2}{n})^{(n+1)/2} } 
\end{align}
The Student-t distribution is widely used in statistics to compare whether two populations have the same mean, given
a priori equal variance.    In the
simplest case where two sets of measurements $x_{1}\ldots x_{N_x}$ and $y_1\ldots y_{N_y}$ have the same sample size, and
$s_x,\bar{x}$ are defined as
\begin{align}
\bar{x} & \equiv \frac{1}{N_x} \sum_k x_k  \\
\label{eq:UnbiasedSigma}
s_x^2 &\equiv \frac{1}{N_x} \sum_k (x_k-\bar{x})^2
\end{align}
and similarly for $s_y,\bar{y}$, 
then the following quantity is $t$ distributed with $N_1+N_2-2$ degrees of freedom:
\begin{eqnarray}
{\cal T} = \frac{\bar{x}-\bar{y}}{\sqrt{ ((N_x-1) s_x^2+(N_y-1)s_y^2)\frac{N_{x}^{-1}+N_y^{-1}}{N_x+N_y-2}}}
\end{eqnarray}
Note that a specific value of ${\cal T}$ corresponds to a difference in means between $x,y$ by of order
$\sigma/\sqrt{N}$.  In our case, with $N\simeq 1000$

The $F_{n_1,n_2}$ distribution arises from the distribution of the ratio 
$F=\frac{y_1/n_1}{y_2/n_2}$ of
two  independent $\chi^2$-distributed random variables  $y_1, y_2$ with $n_1$ and $n_2$ degrees of freedom,
respectively.  The probability distribution of $F$ is
\begin{align}
p(F) &= \frac{\Gamma(\frac{n_1+n_2}{2}) (n_1/n_2)^{n_1/2} F^{\frac{n_1}{2}-1}}{
\Gamma(n_1/2)\Gamma(n_2/2)[
   1+ (n_1 F/n_2)
]^{(n_1+n_2)/2}
}
\end{align}
The  $F$ distribution is most widely used in frequentist hypothesis testing (e.g.,  comparing the residuals after a fit
to an independent estimate of the sample variance).  For the purposes of this study, however, we point out that the
unbiased estimate of the standard deviation [$s_x^2$; Eq. (\ref{eq:UnbiasedSigma})] is proportional to a
$\chi^2$-distributed random variable.  As a result, the suitably-weighted ratio
\begin{eqnarray}
F = s_x^2/s_y^2
\end{eqnarray}
can be used in a standard two-sample F test with $n-1,n-1$ degrees of freedom to assess whether the distributions of $x$
and $y$ have the same variance.  
In practice, because the second moment and therefore the   $F$-test is very sensitive to non-normality, we construct an
\emph{empirical} distribution of $F$, based on two samples known to be drawn from the same distribution: independent repetitions of
the \ST{T2} analysis.

\bibliography{paperexport}

\end{document}